\documentclass[aps,twocolumn,preprintnumbers,footinbib,prl,superscriptaddress,10pt]{revtex4-1}

\makeatletter
\def\l@subsubsection#1#2{}
\def\l@subsubsubsection#1#2{}
\makeatother

\usepackage[utf8]{inputenc}

\usepackage{graphicx,amssymb,amsmath,amsthm,amsfonts,epsfig,epsf}
\usepackage[usenames,dvipsnames,svgnames,table]{xcolor}
\usepackage{epstopdf}
\definecolor{darkred}{rgb}{0.5,0,0}
\usepackage{aas_macros}
\usepackage{bm}
\usepackage{dcolumn}
\usepackage{latexsym}
\usepackage{rotating}
\usepackage{longtable}

\usepackage{enumerate}
\usepackage{tensor,multirow}
\usepackage{url}
\usepackage[linktocpage]{hyperref}

\def\be{\begin{equation}}
\def\ee{\end{equation}}
\newcommand{\beq}{\begin{eqnarray}}
\newcommand{\eeq}{\end{eqnarray}}

\def\ba{\begin{align}}
\def\ea{\end{align}}

\begin{document}
\title{The physics of black hole binaries:\\ geodesics, relaxation modes and energy extraction}

\author{Laura Bernard}
\affiliation{Perimeter Institute for Theoretical Physics, 31 Caroline Street North Waterloo, Ontario N2L 2Y5, Canada}
\affiliation{LUTH, Observatoire de Paris, PSL Research University, CNRS, Universit\'e Paris Diderot, Sorbonne Paris Cit\'e, 5 place Jules Janssen, 92195 Meudon, France}

\author{Vitor Cardoso}
\affiliation{CENTRA, Departamento de F\'{\i}sica, Instituto Superior T\'ecnico -- IST, Universidade de Lisboa -- UL,
Avenida Rovisco Pais 1, 1049 Lisboa, Portugal}
\affiliation{Theoretical Physics Department, CERN 1 Esplanade des Particules, Geneva 23, CH-1211, Switzerland}

\author{Taishi Ikeda}
\affiliation{CENTRA, Departamento de F\'{\i}sica, Instituto Superior T\'ecnico -- IST, Universidade de Lisboa -- UL,
Avenida Rovisco Pais 1, 1049 Lisboa, Portugal}

\author{Miguel Zilh\~ao}
\affiliation{CENTRA, Departamento de F\'{\i}sica, Instituto Superior T\'ecnico -- IST, Universidade de Lisboa -- UL,
Avenida Rovisco Pais 1, 1049 Lisboa, Portugal}

\begin{abstract}
Black holes are the simplest macroscopic objects, and provide unique tests of General Relativity.
They have been compared to the Hydrogen atom in quantum mechanics.
Here, we establish a few facts about the simplest systems bound by gravity: black hole binaries. We provide strong evidence for the existence of ``global'' photosurfaces surrounding the binary, and of binary quasinormal modes leading to exponential decay of massless fields when the binary spacetime is slightly perturbed. These two properties go hand in hand, as they did for isolated black holes.
The binary quasinormal modes have high quality factor and may be prone to resonant excitations.
Finally, we show that energy extraction from binaries is generic and we find evidence of a new mechanism -- akin to the Fermi acceleration process -- whereby the binary transfers energy to its surroundings in a cascading process. The mechanism is conjectured to work when the individual components spin, or are made of compact stars. 
\end{abstract}

\maketitle

% \tableofcontents

% %%%%%%%%%%%%%%%%%%%%%%%%%%%%
% \section{Introduction}
% %%%%%%%%%%%%%%%%%%%%%%%%%%%%
%%%%%%%%%%%%%%%%%%%%%%%%%%%%%%%%%%%%%%%%%%%%%%%%%%%%%%%%%%%%%%%%%%%%%%%%%%%%%
\noindent{\bf{\em I. Introduction.}}
%%%%%%%%%%%%%%%%%%%%%%%%%%%%%%%%%%%%%%%%%%%%%%%%%%%%%%%%%%%%%%%%%%%%%%%%%%%%%
Einstein's theory of General Relativity (GR) is the most accurate known description of
gravity~\cite{Will:2018bme,Will:2018lln,Will:2014kxa,Berti:2015itd}. One of its outstanding predictions
is the existence of black holes (BHs), {\it vacuum} spacetimes defined by an event horizon,
a null one-way surface. Isolated BHs have been studied for decades. They are extremely simple
\footnote{This simplicity is best epitomized in Chandrasekhar's words: ``In my entire scientific life, extending over forty-five years, the most shattering experience has been the realization that an exact solution of Einstein's equations of GR provides the absolutely exact representation of untold numbers of black holes that populate the universe''~\cite{Chandra1}.},
and fully characterized by their mass and angular momentum~\cite{Kerr:1963ud,Newman:1965my,Bekenstein:1996pn,Carter:1997im,Chrusciel:2012jk,Robinson:1975bv,Robinson,Cardoso:2016ryw}. These properties are instrumental to build templates of gravitational wave (GW) signals generated by dynamical BHs, which eventually led to the first direct detection of GWs~\cite{Abbott:2016blz}. The lack of complex multipolar structure of BH geometries is crucial to perform strong-field tests of the theory, for example
through the late time relaxation of BHs, as a superposition of quasinormal modes (QNMs)~\cite{Berti:2005ys,Berti:2009kk,Cardoso:2016ryw,TheLIGOScientific:2016src,Cardoso:2016rao,Cardoso:2017cqb,Cardoso:2019rvt}. 

By contrast, and due to their inherent complexity, BH binaries (BHBs) are less well studied and understood: 
their GW output and dynamics, {\it when in isolation}, is known very well through post-Newtonian expansion techniques at large separations~\cite{Blanchet:2013haa}. In this approach the individual binary components are stationary vacuum BHs, slightly deformed in response to the companion's field. The dynamical behavior of the BHB itself is poorly understood. Such knowledge can in principle be obtained using numerical methods~\cite{Pretorius:2007nq,Cardoso:2014uka}; however, such techniques 
only probe relatively small timescales using finely tuned initial data. In particular, efforts to date focus mostly on purely vacuum spacetimes describing isolated BHBs which have been evolving solely through GW emission, leading to an inspiral and merger, possibly observable by current or future GW detectors. The simulation timescales can be at most of order of a few thousands $GM/c^3$, with $M$ the total spacetime mass. For stellar-mass components, these are of order of one second or less, but BHBs can live for millions of years on tight orbits. Thus, new effects may be triggered and relevant on large timescales. Do perturbed BHBs also have characteristic ringdown modes, and can they be resonantly excited? Do BHBs amplify incoming, low-frequency radiation? Here, we provide a framework for studying these open questions and answer some of them.

%%%%%%%%%%%%%%%%%%%%%%%%%%%%%%%%%%%%%%%%%%%%%%%%%%%%%%%%%%%%%%%%%%%%%%%%%%%%%
\noindent{\bf{\em II. Setup: a black hole binary in post-Newtonian theory.}}
%%%%%%%%%%%%%%%%%%%%%%%%%%%%%%%%%%%%%%%%%%%%%%%%%%%%%%%%%%%%%%%%%%%%%%%%%%%%%
Consider a BHB, whose dynamics are governed by vacuum Einstein equations.
For the reasons outlined (computational expenses, dependence on initial data, etc), instead of numerically-generated spacetimes, we use the approximate BHB spacetime discussed in Ref.~\cite{Mundim:2013vca}. The construction relies on the theory of matched asymptotic expansions and proceeds as follows. The spacetime is divided in three different regions, see Fig.~2 in Ref.~\cite{Mundim:2013vca}. There are two inner zones sufficiently close to each BH, characterized with BH perturbation techniques (the metric perturbation is described by tidal fields generated by the companion; the tidal moments include the quadrupole and octupole deformation and their time derivatives); the inner zone is ``stitched'' to a near zone described by a post-Newtonian expansion (to second post-Newtonian order), itself matched to a far zone which is described using a multipolar, post-Minkowskian formalism. In addition, there are two buffer zones defined as the region of spacetime where the three main zones overlap. The existence of such overlapping regions is crucial to constructing the matched metric. In particular, the multipolar post-Newtonian formalism used to build the near- and far-zone metrics ensures that they are matched by construction~\cite{Blanchet:2013haa}.
Such spacetime was implemented and used to investigate the physics of accretions disks in the presence of BHBs~\cite{Noble:2012xz,Zilhao:2014ida,Bowen:2017oot,Bowen:2016mci}.

\begin{figure}[htb]
\begin{tabular}{c}
\includegraphics[width=8cm]{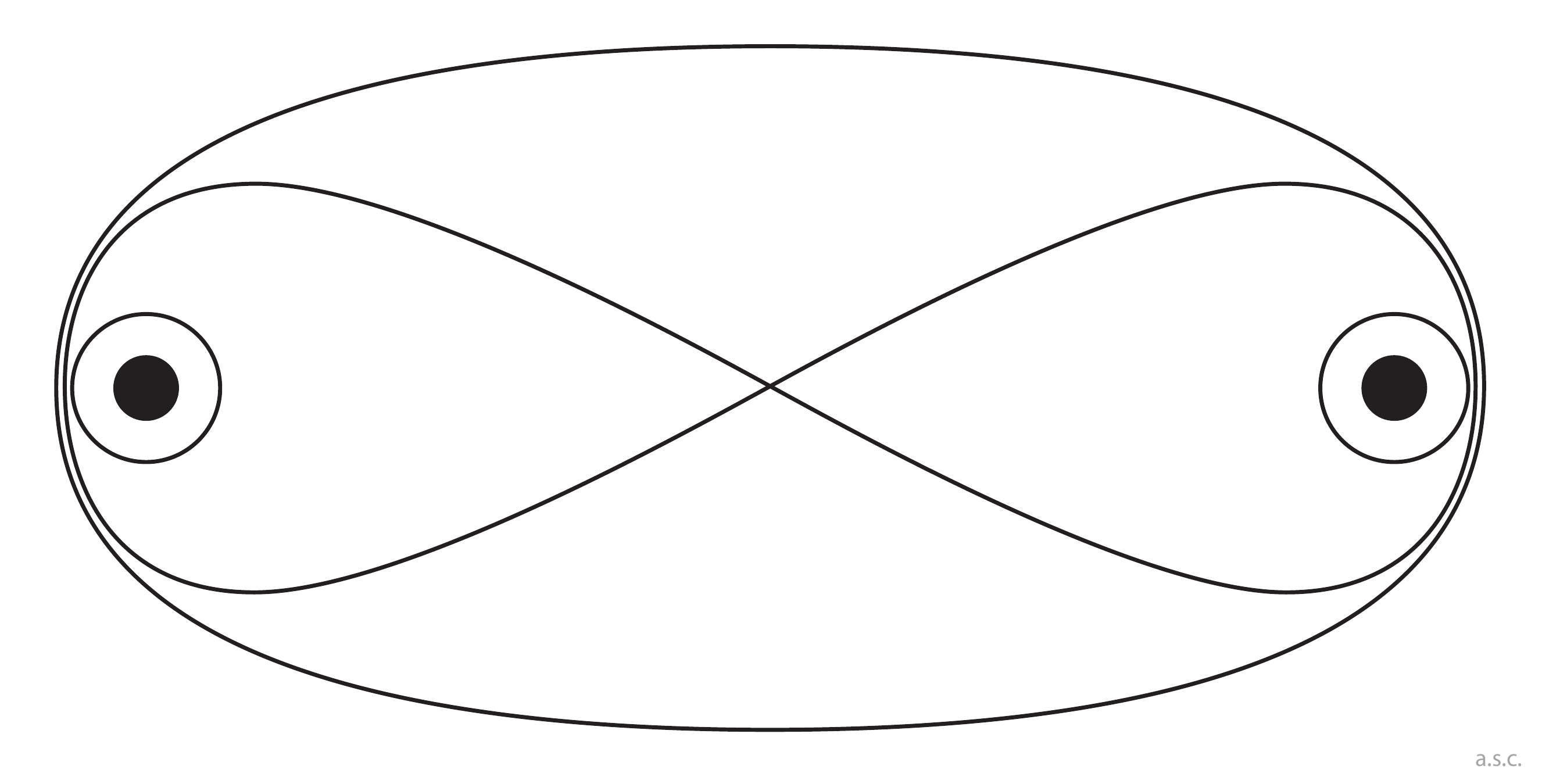}
\end{tabular}
\caption{Cartoon of closed null geodesics in the background of {\it nearly-static}, non-spinning BHBs. Each BH, depicted by a full black circle, is surrounded by its own light ring, lying at $r=3M$ in standard Schwarzschild coordinates, when their separation is large. Two other null closed trajectories are possible: a global non-intersecting null geodesic and an eight-shaped trajectory. Such null geodesics were found in extremal BH or analog spacetimes~\cite{Assumpcao:2018bka,Shipley:2016omi,Chandrasekhar:1989vk}; we also find them in the matched spacetime describing a vacuum BHB. A similar cartoon describes also some timelike geodesics. \label{fig:geodesics} }
\end{figure}
Here onwards we set $G=c=1$, and focus exclusively on equal mass binaries, of total ADM mass $M$ separated by a coordinate distance $L$
(proper separations are very close to the coordinate distances we discuss). 
``Nearly closed'' (i.e., curves of period $2\pi$ to an accuracy which increases with separation $L$ and which self-intersect at least once) null geodesics in such a construction for a vacuum BHB are shown in Fig.~\ref{fig:geodesics}.
We found three types of null geodesics: the ``standard'' null circular geodesic surrounding each BH (which in Schwarzschild coordinates sits at a coordinate distance $\sim r_H/2$ away from the horizon, with $r_H$ the Schwarzschild radius of a single BH), a global non-intersecting geodesic and an eight-shaped trajectory. Such null geodesics were found in toy models in the past, in extremal BH or analog spacetimes~\cite{Assumpcao:2018bka,Shipley:2016omi,Chandrasekhar:1989vk}. The distance of closest approach of the global non-intersecting geodesic to each BH horizon is $ (0.6\pm 0.03) r_H$ and its period $T=2L+31\,(\pm 1)$ for the separations we studied ($L=[20,40]M$). Its distance of closest approach to the center of mass is not strongly sensitive on $L$ and is $y/M= 3.5\pm 0.07$ for these separations. It is natural to speculate that such global geodesic can be identified with the null circular geodesic of the final BH, when such binary merges. Instability of such orbits is clear from their numerical search
(fine-tuning is necessary). A complete picture of geodesic motion in BHB spacetimes is outside the scope of this work.
% we did however search for two other features: Lagrange points and the tightest bound timelike motion.
%We identified the classical stable Lagrange points $L4, L5$~\cite{1994Icar..112..465M,Schnittman:2010br}, at a distance $\sim %L\sqrt{3}/2$ from the center of mass and at $90$ degrees with the binary, as predicted from a Newtonian% %analysis~\cite{1994Icar..112..465M,Schnittman:2010br}. We found global and (seemingly stable) ``eight-shaped'' timelike motion.

%%%%%%%%%%%%%%%%%%%%%%%%%%%%%%%%%%%%%%%%%%%%%%%%%%%%%%%%%%%%%%%%%%%%%%%%%%%%%%%%%%%%%%%%%%%%%%%%%%%%%%%%%%
\noindent{\bf{\em III. Scattering and binary relaxation: individual and global QNMs, and power-law tails.}}
%%%%%%%%%%%%%%%%%%%%%%%%%%%%%%%%%%%%%%%%%%%%%%%%%%%%%%%%%%%%%%%%%%%%%%%%%%%%%%%%%%%%%%%%%%%%%%%%%%%%%%%%%%
The closed null geodesics of isolated BHs correspond to a semi-trapping of massless waves; accordingly, they provide useful
information on the characteristic modes of vibration (QNMs) of BHs~\cite{Ferrari:1984zz,Cardoso:2008bp,Cardoso:2016rao,Cardoso:2017cqb}. It is thus natural to associate the previous global null geodesics with binary relaxation, i.e., with the hitherto unknown QNMs of BHBs. An analytic understanding of these issues for BHBs is challenging, and we turn instead to the numerical simulation of massless scalar fields in the BHB background spacetime described above.

The dynamics of the BHB is governed by vacuum GR, as described previously. We ignore the backreaction of the test scalar field on the BHB spacetime, an approximation which is valid for all realistic setups known to us (except when the scalar field mimics GWs and the BHB is in the last stages of inspiral). Thus, the scalar field is governed by the Klein-Gordon equation $\square \Phi(t,\vec{x})=0$, in a known (albeit time-dependent) background. We use purely ingoing initial data of the form
\begin{eqnarray}
\Phi(0,\vec{x})&\equiv&\Phi_0=\frac{\sin{\omega r}\,W(r)}{r}\,e^{-(r-r_{0})^{2}/\sigma^2 }\,,\label{initial_data}\\
%
%\mathcal{L}_{n}\Phi_0&=&\frac{\Phi_0+x^{i}\partial_{i}\Phi_0}{r}-\frac{\beta^{i}x_{i}x^{j}\partial_{j}\Phi_0}{r^{2}}\,,
%\end{eqnarray}
%\dot{\Phi}_{0}&=&\partial_{r}\Phi_{0}+\frac{\Phi_{0}}{r}\,,
\partial_{t}\Phi(0,\vec{x})&=&\partial_{r}\Phi_{0}+\frac{\Phi_{0}}{r}\,,
\end{eqnarray}
where $r$ is the radial coordinate and $W(r)$ is a window function that smooths the $r=0$ behavior. Here, $r_{0},\,\sigma$ characterize the typical radius and width of the initial ingoing scalar field, while $\omega$ characterizes its frequency.
The initial field amplitude is irrelevant, since the Klein-Gordon equation is linear. To numerically evolve the scalar field, we employ the code presented in Refs.~\cite{Cunha:2017wao,Assumpcao:2018bka}, which makes use of the
\textsc{EinsteinToolkit} infrastructure~\cite{Loffler:2011ay,EinsteinToolkit:web,Zilhao:2013hia} with the
\textsc{Carpet} package~\cite{Schnetter:2003rb,CarpetCode:web}. 

We project the scalar field onto scalar harmonics, $\Phi=\sum_{lm}\Phi_{l,m}Y_{lm}$.
The initial data is spherically symmetric around the center of coordinates, but the presence of the BHB guarantees that upon evolution other components will exist. Note that for symmetry reasons only even modes are excited in the current setup (see also Ref.~\cite{Bentivegna:2008ei}). We studied a variety of different initial parameters and BHB separations. A typical outcome of the evolution is shown in Fig.~\ref{Fig_transient_binaries} (further details are provided in the Appendix). The binary is separated by $L=10M$~\footnote{Such separation is at the limits of applicability of the post-Newtonian approximation~\cite{Mundim:2013vca}; results for larger separations show the same qualitative features.}. Our results show fourth-order convergence (see Appendix).

\begin{figure}[th]
\includegraphics[width=0.5\textwidth]{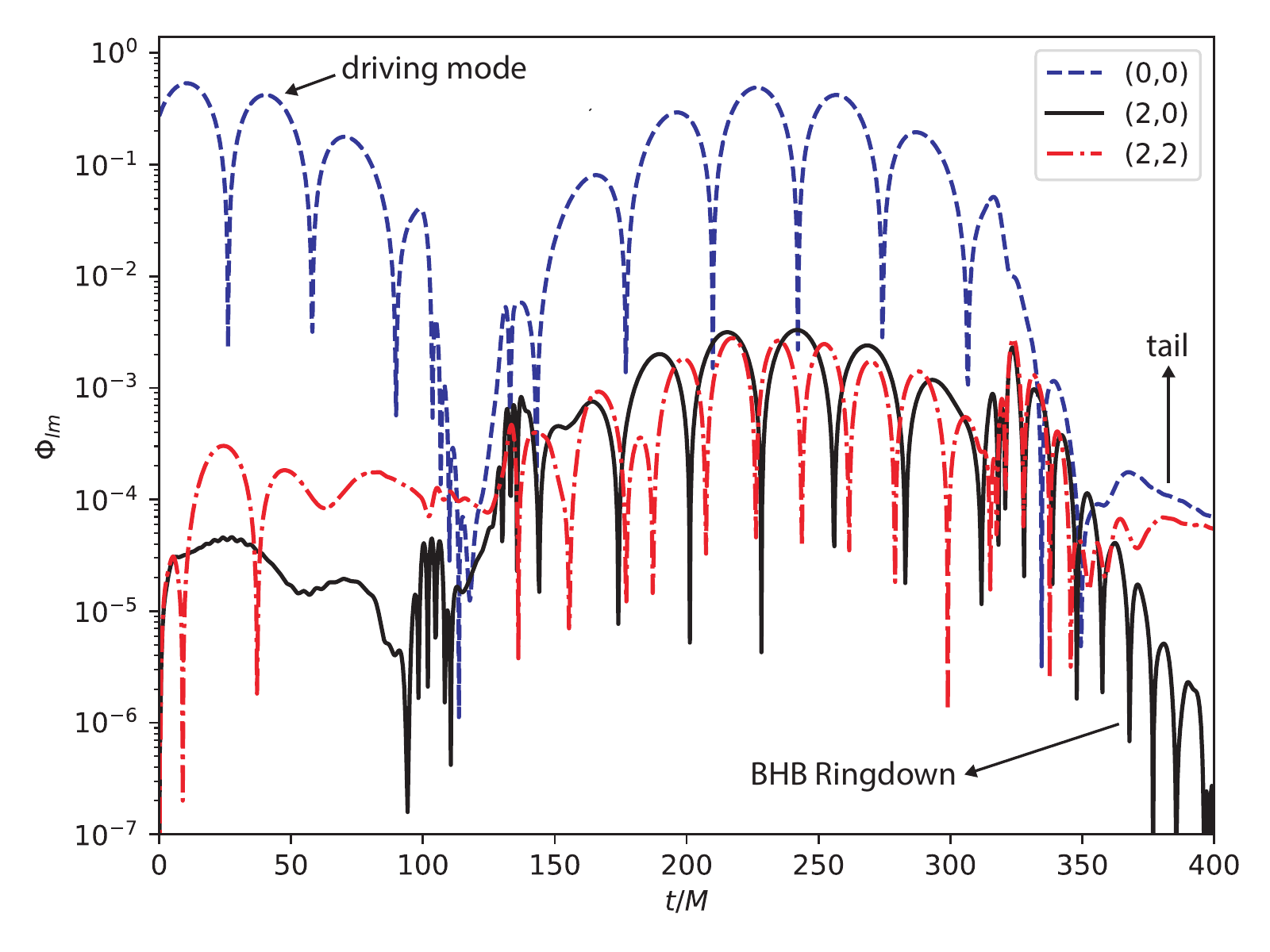}
\caption{Time evolution of an initial spherically symmetric Gaussian pulse, localized at $r_0=100M$, with width $\sigma=40M$, and frequency $M\omega=0.1$. The pulse evolves in a BHB spacetime of separation $L=10M$, % \vc{($12.69M$ proper separation)}. 
the field is extracted at $r=100M$.
\label{Fig_transient_binaries}}
\end{figure}
As seen in Fig.~\ref{Fig_transient_binaries}, the dominant mode is the monopolar one and it drives the dynamics. Initially, the observer sitting at $r=100M$ sees the field heading towards the BHB. At $t\sim 100M$ the observer starts receiving signals which interacted with the BHB. The first clear signal is a strongly damped sinusoid, associated to the ringdown of each {\it individual BH} in the binary. These individual modes are well studied and our results are consistent with theoretical expectations based on linearized calculations~\cite{Berti:2009kk,GRITJHU}. After $t\sim 100M$, the leading monopolar component is now outgoing and drives also higher multipoles.

After the initial driving monopolar mode dies away at $t\sim 320M$, another exponentially damped sinusoid is apparent in the waveform. 
This is one of our main results: BHBs possess global QNMs which describe the ringdown of the binary as a whole. Our results indicate that the ringdown parameters depend only on the mass and separation of the binary and are (to the extent probed by our simulations) independent on the initial data. It is compelling to associate such BHB modes with the closed null geodesics around the BHB, as can be done formally for isolated BHs~\cite{Ferrari:1984zz,Cardoso:2008bp,Cardoso:2016rao,Cardoso:2017cqb}. A reassuring check on the correspondence is that our results are well described by the linear fit
\be
T/M = (1.03\pm 0.04)L/M+8\pm1\,,
\ee
in the range of separations we studied, where we made use of the simulations listed in the Supplement.
The geodesic calculation would have predicted, for $l=m=2$ modes $T\sim L+16)$.
%$T\sim 1/2(2L+T_{\rm LR})$ in the large separation limit. Here $T_{\rm LR}=3\sqrt{3}\pi M\sim 16.3 M$ is the coordinate time a photon takes to go around the photosphere of an isolated, non-spinning BH %of mass $M/2$~\cite{Cardoso:2008bp}. 
In other words, our numerical results are in suggestive agreement with geodesic expectations, lending further support to the association of the global QNMs with global geodesics.
Our results for the decay timescale during this global ringdown phase are less accurate, but suggest that the relaxation timescale $\tau$
also increases with the separation and is of order $10M$ for $L=10M$. 

Notice that the insight from geodesic motion hinges on having two static BHs. This crude approximation is however robust for BHB for which the post-Newtonian expansion is trustworthy: the orbital period scales like $L^{3/2}$ whereas the light travel time scales like $L$. Already for $L=10$ we find an orbital period of $200M$, and a global ringdown timescale of order $10M$. In other words, the binary only traveled a few degrees during the global ringdown stage. The approximation is even better for larger separations. Characteristic vibration modes are generic properties of dissipative systems; a binary system of two stars may also have global QNMs without allowing closed null geodesics~\cite{Kokkotas:1999bd,Cardoso:2019rvt}; although all our results are compatible with such an association, further work is necessary to understand the details and origin of the BHB QNMs.

At very late times, our simulations indicate that the signal dies as a power-law tail in time, $\Phi \sim t^{-\gamma}$. For isolated, non-spinning BH spacetimes and for scalar fluctuations $\gamma=2l+3$~\cite{Ching:1994bd}. Our results indicate $\gamma\sim 7$ for $l=m=0$, which is presumably an indication of mode mixing during the evolution, of the kind seen in isolated but spinning BH geometries~\cite{Zenginoglu:2012us,Burko:2013bra}. 

When the initial wavepacket has a very large width (corresponding to a constant forcing on the BHB), the outgoing pulse is modulated at frequencies $\omega\pm k\Omega$, with $k$ an integer. Such effects were previously seen in the scattering of electromagnetic waves by periodically moving obstacles~\cite{1976IEEEP..64..301V,Jeffery:1980}.

%%%%%%%%%%%%%%%%%%%%%%%%%%%%%%%%%%%%%%%%%%%%%%%%%%%%%%%%%%%%%%%%%%%%%%%%%%%%%%%%%%%%%%%%%%%%%%%%%%%%%%%%%%
\noindent{\bf{\em IV. Energy extraction and instabilities.}}
%%%%%%%%%%%%%%%%%%%%%%%%%%%%%%%%%%%%%%%%%%%%%%%%%%%%%%%%%%%%%%%%%%%%%%%%%%%%%%%%%%%%%%%%%%%%%%%%%%%%%%%%%%
Compact binaries are astrophysical blenders and potential energy sources, either when surrounded by accretion disks or in the context
of fundamental massive fields. Different mechanisms may be associated with energy extraction in the presence of a compact binary:

\noindent {\bf i.} If the individual objects spin, there are ergoregions in the spacetime and each binary component can transfer rotational energy to bosonic fields, through superradiance. Such transfer can be turned into an instability by placing the system inside a cavity~\cite{zeldovich1,zeldovich2,Brito:2015oca}. It is unknown whether binary-intrinsic ergoregions exist (but there are arguments suggesting that superradiance does exist for binaries made of non-spinning objects~\cite{Wong:2019kru}).

\noindent {\bf ii.} A well-known Newtonian energy extraction process, the {\it gravitational slingshot}, transfers kinetic energy from moving planets or stars to scattered probe objects; it is straightforward to show that, within GR, such a mechanism also occurs with light. Light scattering off a BH moving with velocity $v$ can extract (kinetic) energy with an efficiency $\Upsilon \equiv E_{\rm final}/E_{\rm initial}$ up to 
\be
\Upsilon_{\rm max}=\frac{1+v}{1-v}\,.
\ee
The maximum efficiency occurs when the photon scatters with a $180^{\circ}$ angle off an oppositely-moving BH, and is identical to the energy gain of a photon scattering off a moving mirror. For velocities associated to orbital motion in a compact binary, the efficiency can be $1.2$ or higher. We verified such result via explicit scatters of photons off moving BH and BHB geometries~\cite{Thorne:slingshot}. It is conceivable that one photon suffers multiple scatters with the binary (specially if confined).

\noindent {\bf iii.} A binary provides a periodic forcing to external fields; a similar lower-dimensional toy model is known to give rise to instabilities in trapped radiation~\cite{Jeffery:1993}, akin to the Fermi acceleration process of cosmic rays~\cite{Fermi:1949ee}.

These effects or others may all be part of the astrophysics of compact binaries (and hence are all part of realistic simulations, although perhaps not easily identifiable~\cite{vanMeter:2009gu,dAscoli:2018fjw}). We will be interested in {\it confined} binaries, of interest to some dark matter scenarios, where the above mechanisms are expected to trigger instabilities.
Unfortunately, a numerical investigation of these issues using the previous $(3+1)$ setting is challenging; timescales for energy extraction are expected to be very large, and for non-spinning BHBs (the ones we are currently able to simulate in our setup) absorption at the horizon will likely quench or strongly suppress any possible energy extraction mechanism (but see Ref.~\cite{Wong:2019kru}): a BH of mass $M_{\rm BH}$ has an absorption cross section (for scalars) of $20 k \pi M_{\rm BH}^2$~\cite{MTB,Macedo:2013afa,Das:1996we} with $k=27/20, 16/20={\cal O}(1)$ for high and low frequency radiation respectively. Because of this, a naive expectation is that a BHB in a cavity of size $R_{\rm ext}$ will contribute to a decrease in the energy inside that cavity at a rate $dE/dt\sim -\lambda E$, $\lambda\sim 10 k(M/L)^2/R_{\rm ext}$.

\begin{figure}[th]
\includegraphics[width=0.5\textwidth]{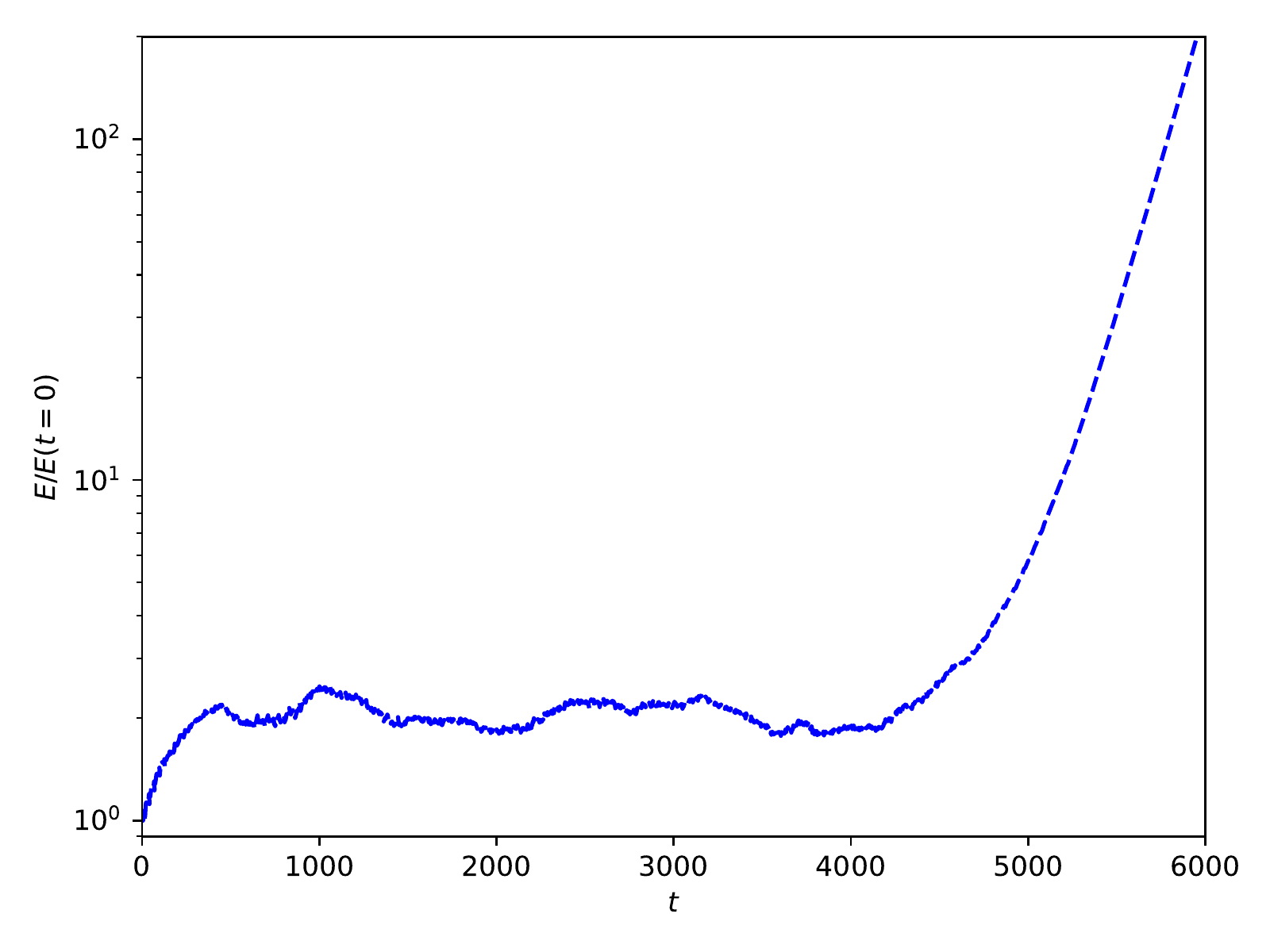}
\caption{Total energy in a scalar field inside a cavity composed of a circular reflecting boundary and two reflecting
circles in circular orbit around each other. The boundary radius is $R_{\rm ext}=30$, each object has a radius $0.5$ and is on a
circular orbit of radius $5$ with angular velocity $\Omega=0.14$. \label{fig:sph_omega014}}
\end{figure}
To emulate compact binaries of spinning BHs or neutron stars, we take a binary of two reflecting objects in flat $(2+1)$-dimensions, and we evolve a massless scalar, with simple gaussian initial profile. This setup allows for a very large number of orbits and reflections in the confining cavity to be evolved. We evolved the system numerically, for different values of orbital frequency $\Omega$, separation $L$ and cavity size $R_{\rm ext}$, with the \textsc{EinsteinToolkit} infrastructure using the code described
in Ref.~\cite{Assumpcao:2018bka}. The total integrated energy inside the cavity is shown in Fig.~\ref{fig:sph_omega014} for one such binary. The total energy {\it increases} with time, and at late times such increase is exponential. Our results indicate that this growth happens only when the orbital frequency is of the order of the light travel time inside the cavity. Although a toy model, this is probably the first example of instabilities triggered by binaries. There is nothing intrinsic to lower dimensional spacetimes; the arguments above suggest that it may have a counterpart in (3+1) setups as well. Enclosed BHBs are currently being studied~\cite{Bernard_follow}.

%%%%%%%%%%%%%%%%%%%%%%%%%%%%%%%%%%%%%%%%%%%%%%%%%%%%%%%%%%%%%%%%%%%%%%%%%%%%%
\noindent{\bf{\em V. Discussion.}}
%%%%%%%%%%%%%%%%%%%%%%%%%%%%%%%%%%%%%%%%%%%%%%%%%%%%%%%%%%%%%%%%%%%%%%%%%%%%%
The role and simplicity of BHs in GR has often been compared to the Hydrogen
atom in the development on quantum mechanics. It is compelling to draw a
parallel between BH binaries and the simplest possible molecule, that of the
Hydrogen molecule ion~\cite{Burrau:1927a,Wilson:1928,Hylleraas:1931,Leaver:1986}.
The null geodesics around isolated BHs have a counterpart in wave dynamics, as QNMs of the geometry. In a
quantum-mechanical picture they would correspond to the bound states of the
Hydrogen atom. Likewise, the global geodesics for BHBs (see
Fig.~\ref{fig:geodesics}) may be tightly connected to global QNMs
(Fig.~\ref{Fig_transient_binaries}), the analog of molecular bound states; it is amusing to note that such correspondence can be taken a step further for
an exact BHB solution, the Majumdar Papapetrou geometry describing a pair
of extremal BHs; the Klein-Gordon equation in this background assumes the same form as the Schr\"{o}dinger equation
describing a positron in the ionized Hydrogen molecule (see Appendix). 
The successful, effective-one-body treatment of post-Newtonian theory for the BHB problem, uses very explicit hydrogen-atom analogies to construct an inspiraling waveform model~\cite{Buonanno:1998gg}.
It is tempting to expect that such methods can also yield insight on the problem of BHB relaxation.
A quantitative correspondence can be established between geodesics and QNMs of isolated BHs; a
similar analysis for BHBs is missing, but would be fundamental for the program of understanding
the high frequency QNM regime and its possible connection to null geodesics.
BH spectroscopy makes use of the QNMs of isolated BHs to infer their mass and spin~\cite{Berti:2005ys,Berti:2007zu,Berti:2016lat}. 
Perhaps BHB spectroscopy is within reach, where their separation can also be estimated from the global modes. The excitation mechanism of BHB modes could happen, for example, via three-body interactions.

It is a fact that, since the QNMs of isolated BHs are associated to photospheres -- and hence to the largest frequency in such spacetime -- the resonant excitation of such modes is impossible in astrophysical setups.
However, the new global modes associated with the entire BHB geometry are associated to a new scale (the binary separation), and have lower frequency than each of the individual BH QNMs. It is therefore conceivable that a particle orbiting one of the BHs can resonantly excite the global modes. The condition for this to happen is that the orbital period of the small particle equals the period of the QNMs. We find that the particle orbiting at radius $r_p$ around one of the BHs should satisfy $r_p/M_{\rm BH}=0.466(L/M+3\pi\sqrt{3}/2)^{2/3}$ ($M_{\rm BH}$ is the mass of each individual BH). For $L=38M$ for example, resonant excitation is possible when the particle reaches the innermost stable circular orbit of one individual BH. The consequence could be enhanced emission of GWs from the binary.

There are compelling arguments suggesting that compact binaries may be prone to energy transfer to other degrees of freedom
(most notably to fundamental scalars, for example) when inside a cavity. This artificial setup mimics well physically motivated scenarios consisting on massive degrees of freedom, such as massive scalars or vectors, which are proxies or serious candidates for dark matter (for instance when dealing with axion-like particles)~\cite{Brito:2015oca}. Our results strongly suggest that a new type of instability may be active, potentially interesting to improve constraints on dark matter models. The mechanism resembles a parametric instability~\cite{Yang:2014tla}, but we find complex growth patterns in the field. The instability growth rate is larger for larger angular velocities, thus a natural question remains open, which 
our toy model is unable to answer: is the instability relevant for an astrophysical binary, driven by GW emission?
In other words, is there any regime during which the growth rate is important during a binary lifetime?~\cite{Huan}
%One might argue that a binary of compact stars enclosed in a cavity is an effective way of forcing an enclosed field to sweep the %entire phase space, and to increase the entropy. It is therefore tempting to conjecture that -- once backreaction on the spacetime is %included -- BH production might occur.

%%%%%%%%%%%%%%%%%%%%%%%%%%%%%%%%%%%%%%%%%%%%%%%%%%%%%%%%%%%%%%%%%%%%%%%%%%%%%
\noindent{\bf{\em Acknowledgments.}}
%%%%%%%%%%%%%%%%%%%%%%%%%%%%%%%%%%%%%%%%%%%%%%%%%%%%%%%%%%%%%%%%%%%%%%%%%%%%%
%
We thank Ana Carvalho, for useful input and for producing some of the figures for us.
We thank David Hilditch for useful advice regarding the numerical implementation. We are indebted to Lorenzo Annulli, Emanuele Berti, Kfir Blum, Miguel Correia, Will East, Pedro Ferreira, Masashi Kimura, Luis Lehner, Jos\'e Nat\'ario, Rodrigo Vicente, Leong Khim Wong and Huan Yang for useful comments on a version of this manuscript, and for discussions on several issues related to this work.
T.I.\ and M.Z.\ are indebted to the Theory Division of CERN for the warm hospitality.
M.Z.\ also thanks the hospitality of Perimeter Institute for Theoretical
Physics, where part of this work was done.
V.C.\ acknowledges financial support provided under the European Union's H2020 ERC 
Consolidator Grant ``Matter and strong-field gravity: New frontiers in Einstein's 
theory'' grant agreement no. MaGRaTh--646597.
M.Z.\ acknowledges financial support provided by FCT/Portugal through the IF
programme, grant IF/00729/2015.
This research was supported in part by Perimeter Institute for Theoretical Physics. Research at Perimeter Institute is supported by the Government of Canada through the Department of Innovation, Science and Economic Development Canada and by the Province of Ontario through the Ministry of Economic Development, Job Creation and Trade.
This project has received funding from the European Union's Horizon 2020 research and innovation programme under the Marie Sklodowska-Curie grant agreement No 690904.
We acknowledge financial support provided by FCT/Portugal through grant PTDC/MAT-APL/30043/2017.
The authors would like to acknowledge networking support by the GWverse COST Action 
CA16104, ``Black holes, gravitational waves and fundamental physics.''
Computations were performed on the ``Baltasar Sete-Sois'' cluster at IST and XC40 at YITP in Kyoto University.
%
% \end{acknowledgments}
%%%%%%%%%%%%%%%%%%%%%%%%%%%%%%%%%%%%%%%%%%%%%%%%%%%%%%%%%%%%%%%%%%%%%%%%%%%%%

%%%%%%%%%%%%%%%%%%%%%%%%%%%%%%%%%
\appendix 

%%%%%%%%%%%%%%%%%%%%%%%%%%%%%%%%%%%%%%%%%%%%%%%%%%%%%%%%%
\section{Simulations, convergence and radial dependence}
%%%%%%%%%%%%%%%%%%%%%%%%%%%%%%%%%%%%%%%%%%%%%%%%%%%%%%%%%
Table~\ref{Table of simulations} summarizes the parameters of our simulations, for the scattering of gaussian wavepackets off BHBs.
\begin{table}[ht]
 \begin{tabular}{|c||c|c|c|c|c|c|c|c|c|c|}
\hline
Name&$L$&$\sigma$&$\omega$&$r_{0}$&$T_{20}$\\%&$\varphi$&$r_{\rm in}$&$r_{\rm out}$&$w$&$n$\\
\hline\hline
BHB1&$20$&$40$&$0.1$&$100$&26.6\\%&$0$&$20$&$200$&$10$&$5$\\
\hline
BHB2&$20$&$40$&$0.05$&$100$&28.0\\%&$0$&$20$&$200$&$10$&$5$\\
\hline
BHB3&$10$&$40$&$0.1$&$100$&19.0\\%&$0$&$20$&$200$&$10$&$5$\\
\hline
BHB4&$40$&$40$&$0.1$&$100$&49.6\\%&$0$&$20$&$200$&$10$&$5$\\
\hline
BHB5&$20$&$40$&$0.01$&$100$&27.2\\%&$0$&$20$&$200$&$10$&$5$\\
\hline
BHB6&$20$&$40$&$0.02$&$100$&27.0\\%&$0$&$20$&$200$&$10$&$5$\\
\hline
BHB7&$20$&$80$&$0.2$&$100$&28.6\\%&$0$&$20$&$200$&$10$&$5$\\
\hline
BHB8&$20$&$80$&$0.5$&$100$&*\\%&$0$&$20$&$200$&$10$&$5$\\
\hline
BHB9&$10$&$4$&$0.1$&$100$&18.6\\%&$0$&$20$&$200$&$10$&$5$\\
\hline
\end{tabular}
\caption{Summary of our simulations. The parameters specify the initial conditions, as in Eq.~\eqref{initial_data} in the main text. 
Here, the mass of each BH is fixed at $0.5$. $T_{20}$ is the period of oscillations of the $l=2, m=0$ component at late-times. Our results show that the $l=m=2$ has a similar period.  The oscillation period $T_{20}$ is extracted using two periods of the late-time oscillation, when available.
For the BHB8 run, the late-time behavior of $T_{20}$ is highly modulated and hard to extract any characteristic frequency.\label{Table of simulations}}
\end{table}
%
%%%%%%%%%%%%%%%%%%%%%%%%%%%%%%%%%%%%%%%%%%%%%%%%%%%%%%%%%%%%%%%%%%%
\subsection{Numerical procedure and convergence analysis}
%%%%%%%%%%%%%%%%%%%%%%%%%%%%%%%%%%%%%%%%%%%%%%%%%%%%%%%%%%%%%%%%%%%
To numerically evolve the scalar field equations in our prescribed metric
background we employ the code presented in
Refs.~\cite{Cunha:2017wao,Assumpcao:2018bka}, which makes use of the
\textsc{EinsteinToolkit}
infrastructure~\cite{Loffler:2011ay,EinsteinToolkit:web,Zilhao:2013hia} with the
\textsc{Carpet} package~\cite{Schnetter:2003rb,CarpetCode:web} for
mesh-refinement capabilities.
We employ the method-of-lines, where spatial derivatives are approximated by
fourth-order finite difference stencils, and we use the fourth-order
Runge-Kutta scheme for the time integration.  Kreiss-Oliger dissipation is
applied to evolved quantities in order to damp high-frequency noise.

\begin{figure}[htbp]
  \centering
  \includegraphics[width=0.45\textwidth]{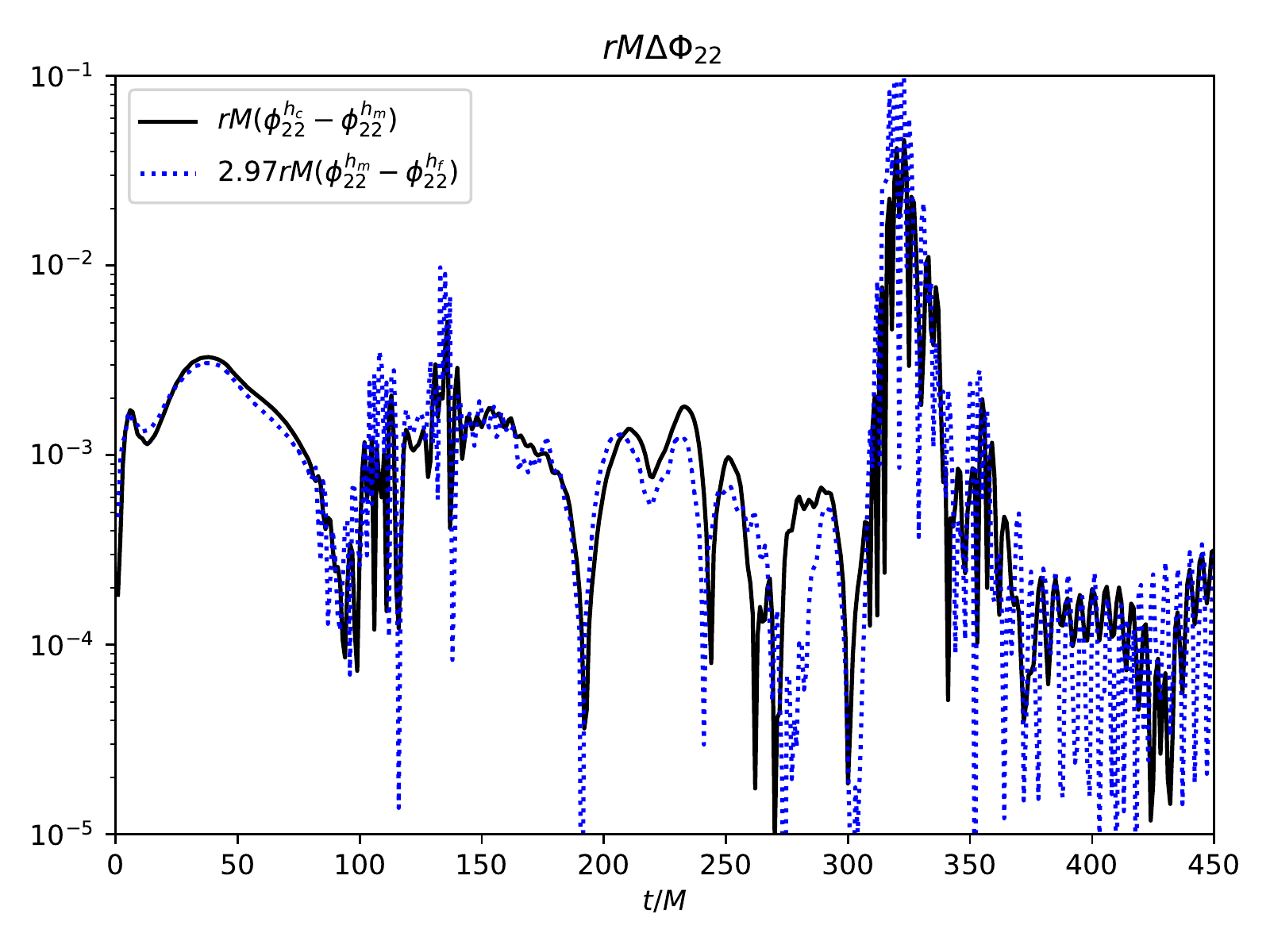}
  \caption[]{Convergence analysis of the $l=2$, $m=2$ multipole of $\Phi$
    extracted at $r=100~M$ for the configuration of
    Fig.~\ref{Fig_transient_binaries} in main text (BHB3), showing good agreement with fourth-order
    convergence. \label{Fig_convergence}}
\end{figure}
Our simulations use finite differencing techniques, which approximate the
continuum solution of the problem with an error that depends polynomially on the
grid spacing $h$,
\begin{equation}
  f=f_{h}+ O\left(h^{n}\right) \,,
\end{equation}
where $n$ is the convergence order.
Since the code we employ uses both second- and fourth-order techniques we expect
this to be reflected in the convergence properties of our results. Consistency
can be checked by evolving the same configuration with coarse, medium and fine
resolution $h_{c}$, $h_m$ and $h_f$. One can then compute the convergence factor
given by
\begin{equation}
\label{eq:Qconv}
Q \equiv \frac{f_{h_c}-f_{h_m}}{f_{h_m}-f_{h_f}} = \frac{h_c^n-h_m^n}{h_m^n-h_f^n} \,.
\end{equation}
To check the convergence of the extracted waveforms we have evolved the
configuration of Fig.~\ref{Fig_transient_binaries} in the main text, with resolutions
$h_{c}= 1.6~M$, $h_{m} = 1.28~M$ and $h_{f} = 1.0~M$ (where this refers to
resolution of the outermost refinement level); the corresponding results are
shown in Fig~\ref{Fig_convergence} for the $l=2$, $m=2$ multipole of $\Phi$. We
have amplified the differences between the medium and fine resolution runs by
the factor 2.97 expected for fourth-order convergence.

%%%%%%%%%%%%%%%%%%%%%%%%%%%%%%%%%%%%%%%%%%%%%%%%%%%%%%%%%%%%%%%%%%%%%%%%%%%%%%%%%%%%%%%%%
\subsection{Dependence on extraction radii}
%%%%%%%%%%%%%%%%%%%%%%%%%%%%%%%%%%%%%%%%%%%%%%%%%%%%%%%%%%%%%%%%%%%%%%%%%%%%%%%%%%%%%%%%
%
\begin{figure}[htbp]
  \centering
\includegraphics[width=0.45\textwidth]{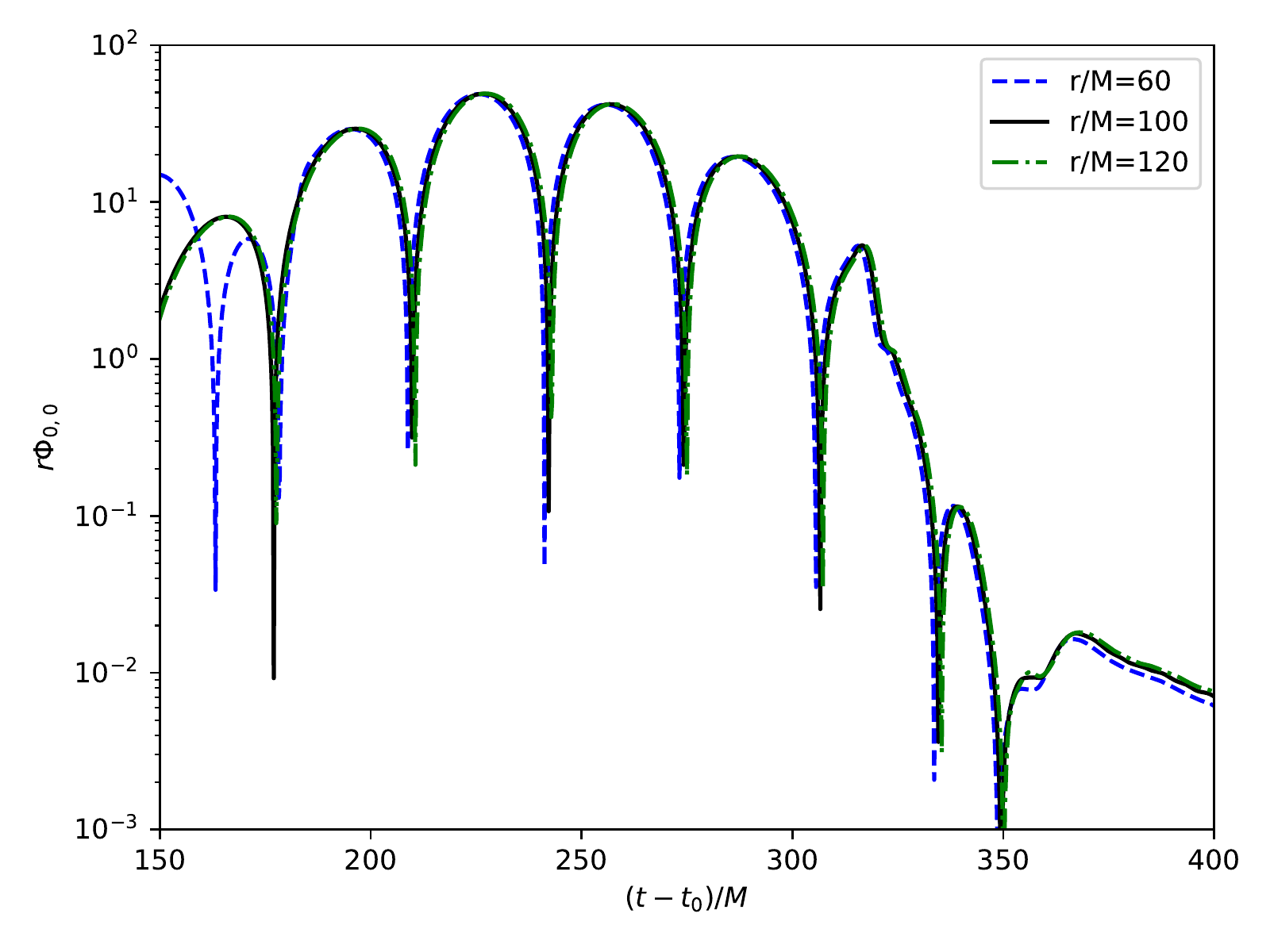}
\includegraphics[width=0.45\textwidth]{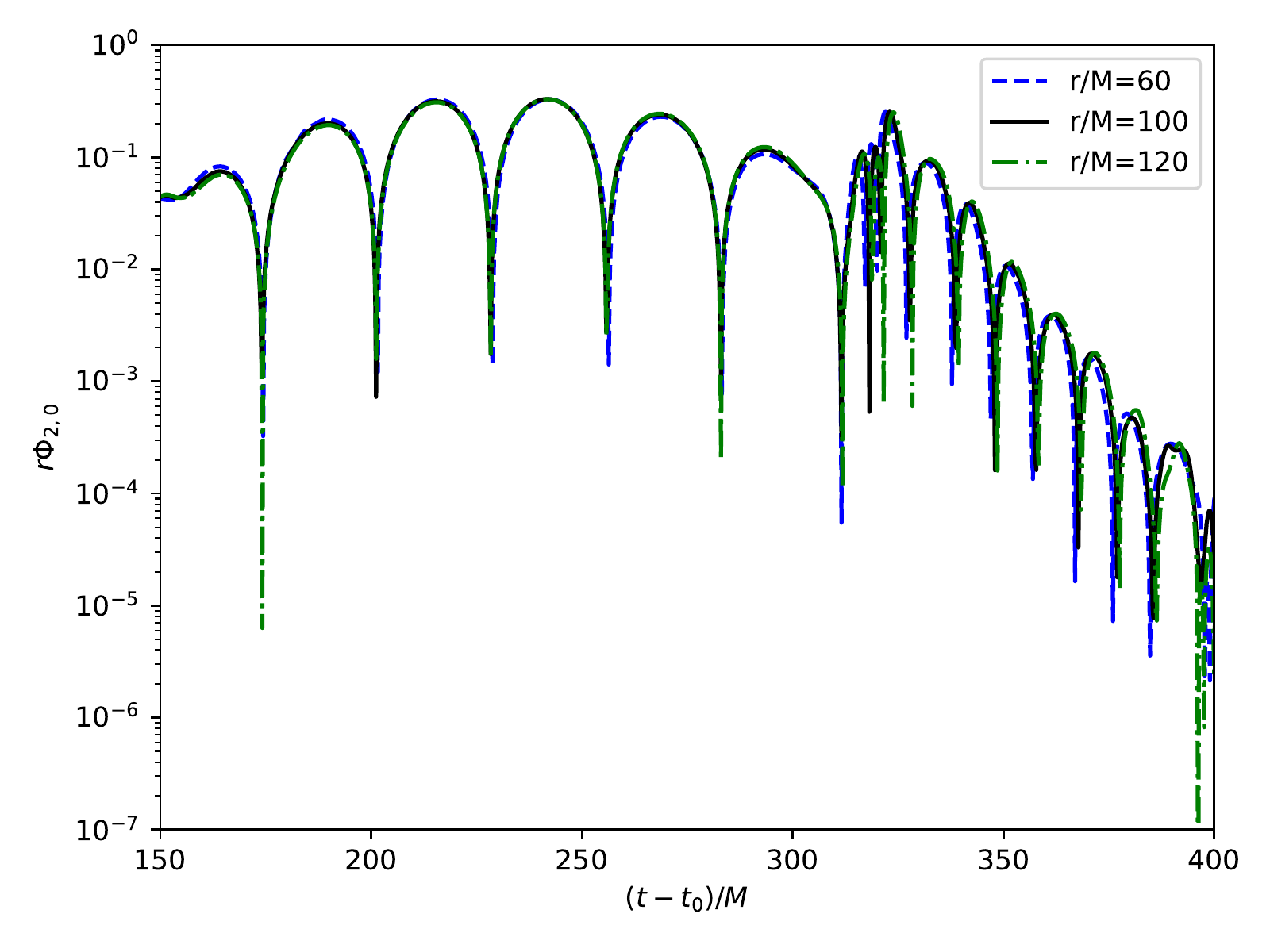}
\caption[]{Radial dependence of the waveform shown in Fig.~\ref{Fig_transient_binaries} of the main text (BHB3) for the $l=m=0$ (top) and $l=2, m=0$ (bottom) modes. Here, we aligned the waveform extracted at three different radii, by properly subtracting the light travel time. As we can see there's a very good overlap of the waveform for the $l=2$ mode, indicating that it is purely outgoing and that one is indeed in the wave zone already and capturing the leading order dependence of the waveform. The driving $l=0$ mode, on the other hand, is not aligned
at early times, showing that indeed it is initially ingoing. \label{Fig_radial_dependence}}
\end{figure}
Figure~\ref{Fig_radial_dependence} shows the waveform extracted at different radii, aligned by their maxima (in other words, aligned by the light travel-time propagation delay).
The consistent overlap between the three signals indicates that the signal is indeed being measured in the wave zone, and that there is no finite-extraction radius artifact.
Notice that the early-time response for the monopole does not align: the reason is that the initial pulse is ingoing. The late time behavior of the monopolar component
on the other hand is perfectly aligned, showing that the pulse is outgoing at these late stages.
%%%%%%%%%%%%%%%%%%%%%%%%%%%%%%%%%%%%%%%%%%%%%%%%%%%%%%%%%%%%%%%%%%%%%%%%%%%%%%%%%%%%%%%%%
\subsection{Dependence on initial data}
%%%%%%%%%%%%%%%%%%%%%%%%%%%%%%%%%%%%%%%%%%%%%%%%%%%%%%%%%%%%%%%%%%%%%%%%%%%%%%%%%%%%%%%%%
%
\begin{figure}[htbp]
  \centering
\includegraphics[width=0.45\textwidth]{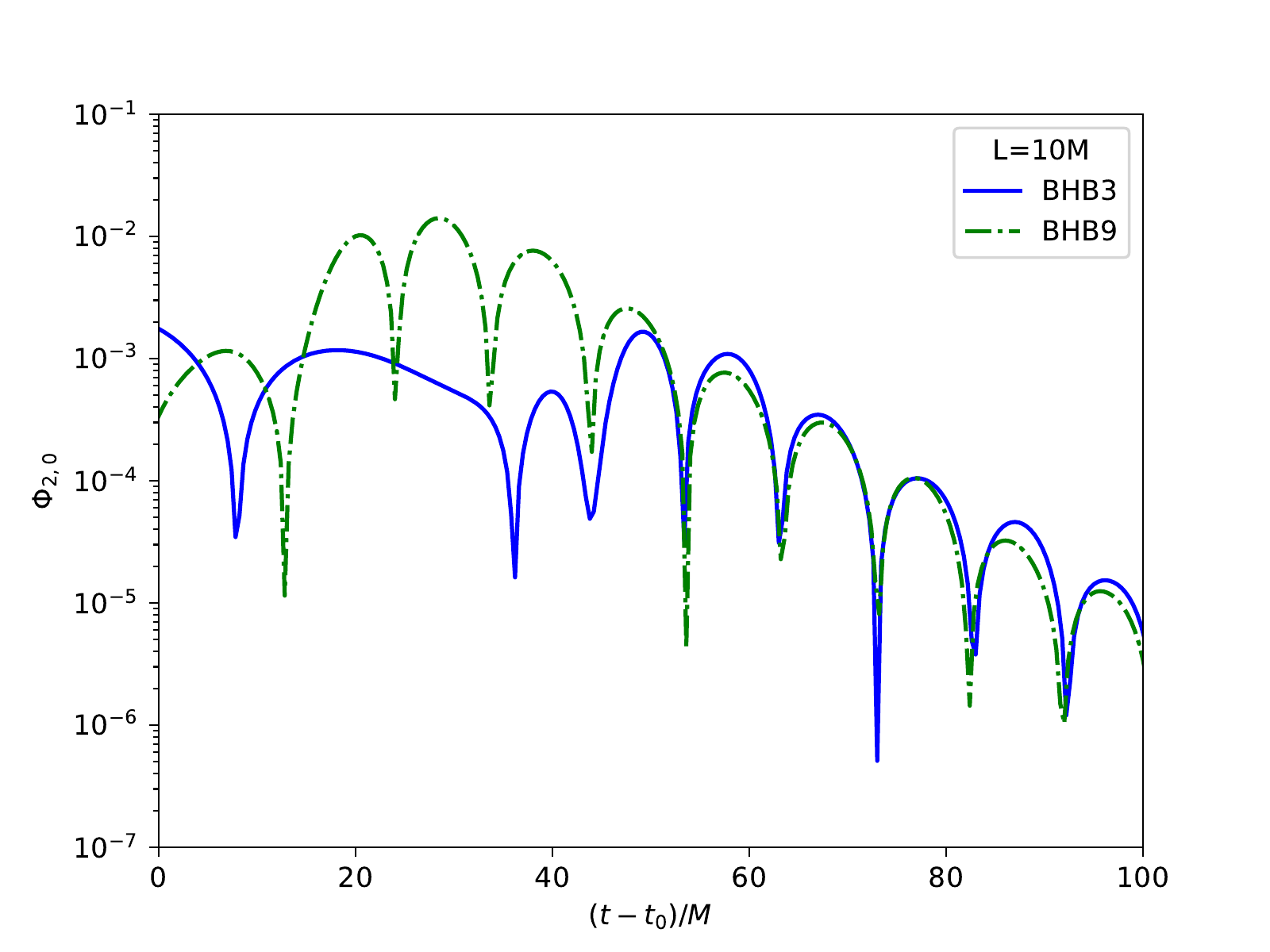} 
\includegraphics[width=0.45\textwidth]{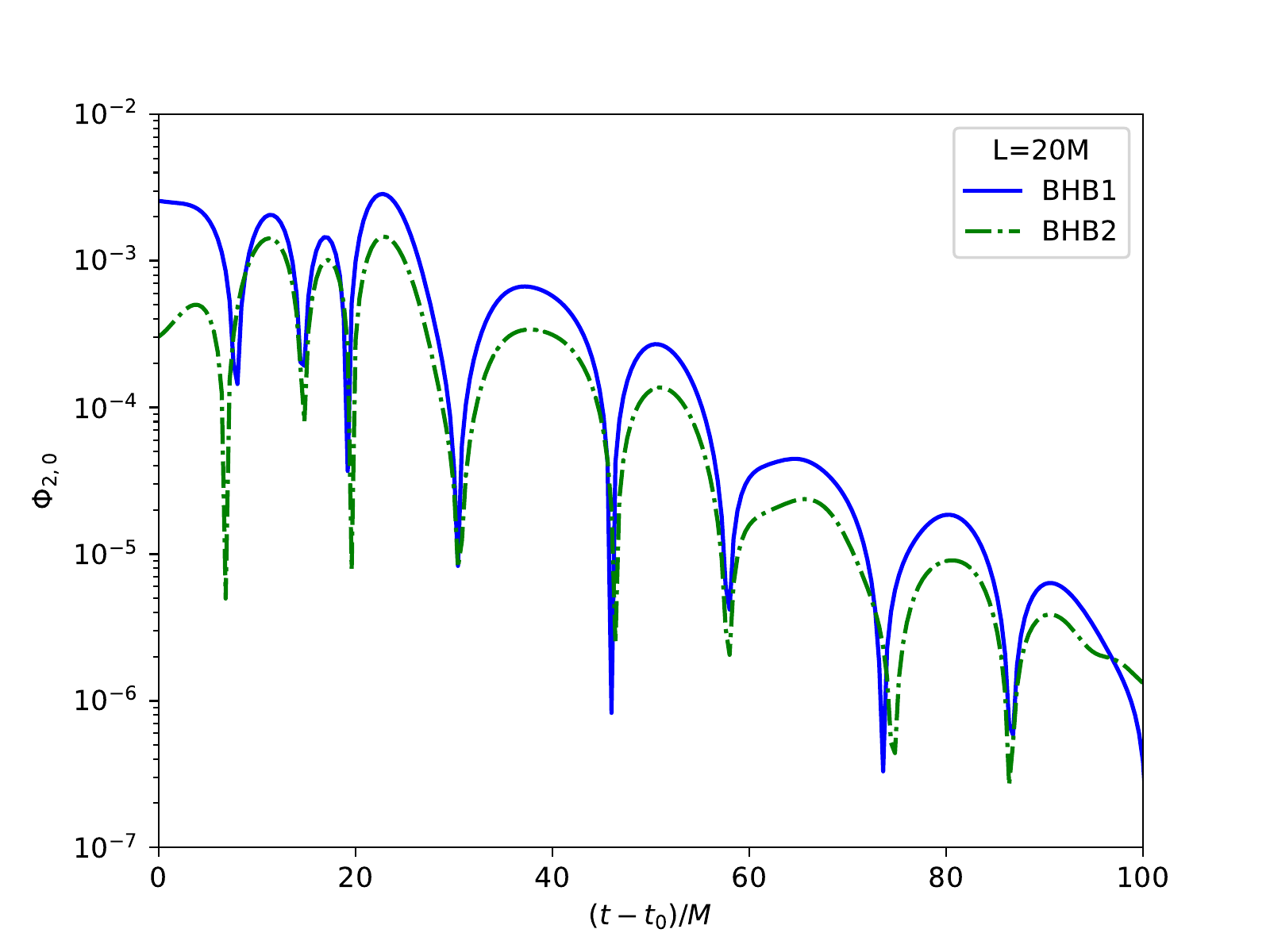}
\caption[]{Waveforms for different initial data, following the notation of Table~\ref{Table of simulations} and extracted at $r=100M$. The waveforms were aligned in time, such that the universality of the ringdown phase is clear.
\label{Fig_Initial_data}}
\end{figure}
Figure~\ref{Fig_Initial_data} shows the waveforms for different types of initial conditions, following the definitions in Table~\ref{Table of simulations}.
The waveforms were aligned in time, to show a clear universal ringdown, the most important result of our work. The ringdown frequency and damping timescale is the same for different initial
conditions, and depends only on the binary parameters (mass and separation). The results also indicate clearly that, although the period and damping timescale scale with separation $L$,
the quality factor seems to be scale independent.

The initial pulse is spherically symmetric. However, on very short timescales the signal develops a quadrupolar ($l=2$) component as well.
This behavior is expected, since one is specifying spherically symmetric initial conditions on
a non-symmetric background. Thus, the field will very quickly sense the non-symmetric background metric. It is possible to show that such non-symmetric component
is weaker at larger distances.
%%%%%%%%%%%%%%%%%%%%%%%%%%%%%%%%%%%%%%%%%%%%%%%%%%%%%%%%%%%%%%%%%%%%%%%%%%%%%%%%%%%%%%%%%
\subsection{Power-law tails}
%%%%%%%%%%%%%%%%%%%%%%%%%%%%%%%%%%%%%%%%%%%%%%%%%%%%%%%%%%%%%%%%%%%%%%%%%%%%%%%%%%%%%%%%
\begin{figure}[htbp]
  \centering
  \includegraphics[width=0.45\textwidth]{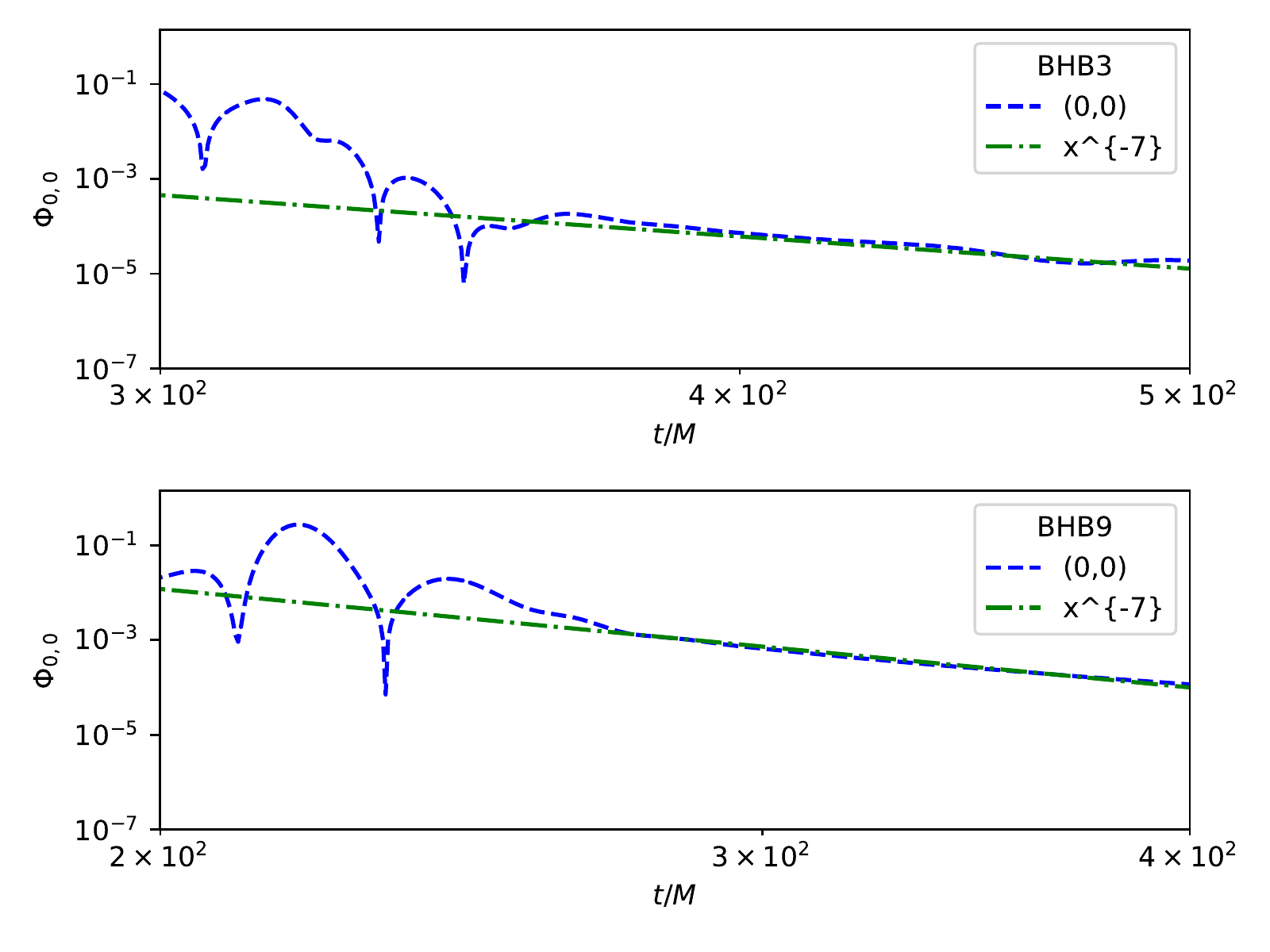}
  \caption[]{Late-time behavior of the driving $l=m=0$ mode for two different simulations. The results are compatible with a power-law tail at late times. \label{Fig_tails}}
\end{figure}
Figure~\ref{Fig_tails} zooms-in on the late-time behavior of the dominant, driving $l=m=0$ simulations for the scattering of gaussian wavepackets.
Our results are consistent with a power-law.

%%%%%%%%%%%%%%%%%%%%%%%%%%%%%%%%%%%%%%%%%%%%%%%%%%%%%%%%%%%%%%%%%%%%%%%%%%%%%%%%%%%%%%%%%
\section{Majumdar-Papapetrou and the di-Hydrogen ionized molecule}
%%%%%%%%%%%%%%%%%%%%%%%%%%%%%%%%%%%%%%%%%%%%%%%%%%%%%%%%%%%%%%%%%%%%%%%%%%%%%%%%%%%%%%%%
There is an exact solution in General Relativity describing two or more static BHs. Such a solution is known as the Majumdar-Papapetrou (MP) solution~\cite{Majumdar:1947eu,Papapetrou:1947a,Hartle:1972ya}. In a cylindrical coordinate system the BHB version of the MP solution is written as
\be
\label{eqn:MPmetric}
ds^2=-\frac{dt^2}{U^2}+U^2\left(d\rho^2+\rho^2d\phi^2+dz^2\right)\,,
\ee
with 
\be
\label{eqn:MP_U}
U(\rho,\,z)=1+\frac{M}{\sqrt{\rho^2+(z-a)^2}}+\frac{M}{\sqrt{\rho^2+(z+a)^2}}\,.
\ee
This solution represents two maximally charged BHs in equilibrium, each with mass $M$ and charge $Q=M$. In these coordinates, their horizons are shrunk to two points at $z= \pm a$ (hence, the parameter $a$ measures the distance between them). The spacetime ADM mass is $2M$.

Some geodesic properties of these solutions were studied before~\cite{Chandrasekhar:1989vk,Assumpcao:2018bka,Shipley:2016omi}.
We will now show the rather remarkable result that the Klein-Gordon equation separates.
Change to planar ``prolate confocal elliptical'' coordinates $\chi, \eta$ (keeping the time and azimuthal coordinates) defined as
\beq
r_1^2\equiv \rho^2+(a-z)^2=a^2(\chi+\eta)^2\,,\\
r_2^2\equiv \rho^2+(a+z)^2=a^2(\chi-\eta)^2\,.
\eeq
The variable $\chi$ plays a role similar to $r$ in standard spherical coordinates, while $\eta$
plays the role of $\cos\theta$. The domains of these variables are $-1\leq\eta\leq 1,\,1\leq\chi\leq \infty$.
In these coordinates the Klein-Gordon equation for $\Psi e^{im\phi-i\omega t}$ can be written as
\beq
0&=&\partial_{\chi}\left((\chi^2-1)\partial_\chi \Psi\right)-\partial_{\eta}\left((\eta^2-1)\partial_\eta \Psi\right)\nonumber\\
&+&\left(\frac{m^2}{\eta^2-1}-\frac{m^2}{\chi^2-1}\right)\Psi+\frac{\omega^2\left(a\chi^2-a\eta^2+2M\chi\right)^4\Psi}{a^2(\chi^2-\eta^2)^3}\nonumber\,.
\eeq
When $a/M\gg 1$, we find that the above is separable and reduces to
\beq
&&\partial_{\chi}\left((\chi^2-1)\partial_\chi \Psi\right)-\partial_{\eta}\left((\eta^2-1)\partial_\eta \Psi\right)\nonumber\\
&+&\left(\frac{m^2}{\eta^2-1}-\frac{m^2}{\chi^2-1}\right)\Psi+a^2\omega^2\left(\chi^2-\eta^2+\frac{8\chi\,M}{a}\right)\Psi=0\nonumber\,.
\eeq
With the ansatz $\Psi=S(\eta)R(\chi)$, we find finally
\beq
&&\partial_{\eta}\left((1-\eta^2)\partial_\eta S\right)+\left(-a^2\omega^2\eta^2-\frac{m^2}{1-\eta^2}+\Lambda\right)S=0\nonumber\,,\\
&&\partial_{\chi}\left((\chi^2-1)\partial_\chi R\right)\nonumber\\
&+&\left(a^2\omega^2\chi^2+8Ma\chi\,\omega^2-\frac{m^2}{\chi^2-1}-\Lambda\right)R=0\,,
\eeq
where $\Lambda$ is a separation constant.
This same system describes the Schrodinger equation (for a positron) in the ionized Hydrogen molecule~\cite{Burrau:1927a,Wilson:1928}.
We thus have a formal equivalence between two similar systems, that of a molecule governed by electromagnetism and 
a simple binary system in full General Relativity. 
The effective-one-body treatment of post-Newtonian theory for the BHB problem, uses very explicit hydrogen-atom analogies to construct an inspiraling waveform model~\cite{Buonanno:1998gg}.
Thus, it is interesting that the converse (i.e., recovering the dynamics of a molecule in quantum mechanics) is borne out of an exact solution in General Relativity.

%%%%%%%%%%%%%%%%%%%%%%%%%%%%%%
\bibliographystyle{apsrev4}
\bibliography{References}

\end{document}